\documentclass[conference]{IEEEtran}
\IEEEoverridecommandlockouts

\usepackage{tikz}
\usepackage{textcomp}
\usepackage{hyperref}
\usepackage{lipsum}

\newcommand\copyrighttext{%
  \footnotesize \textcopyright 2019 IEEE. Personal use of this material is permitted.
  Permission from IEEE must be obtained for all other uses, in any current or future
  media, including reprinting/republishing this material for advertising or promotional
  purposes, creating new collective works, for resale or redistribution to servers or
  lists, or reuse of any copyrighted component of this work in other works.
  DOI: \href{https://doi.org/10.1109/BigData47090.2019.9005504}{10.1109/BigData47090.2019.9005504}}
\newcommand\copyrightnotice{%
\begin{tikzpicture}[remember picture,overlay]
\node[anchor=south,yshift=10pt] at (current page.south) {\fbox{\parbox{\dimexpr\textwidth-\fboxsep-\fboxrule\relax}{\copyrighttext}}};
\end{tikzpicture}%
}

\usepackage{eso-pic}
\usepackage{cite}
\usepackage{float}
\usepackage{amsmath,amssymb,amsfonts}
\usepackage{hyperref}
\usepackage{listings}
\usepackage{algorithmic}
\usepackage{graphicx}
\usepackage{textcomp}
\usepackage{xcolor}
\usepackage{array}
\usepackage{booktabs}
\def\BibTeX{{\rm B\kern-.05em{\sc i\kern-.025em b}\kern-.08em
    T\kern-.1667em\lower.7ex\hbox{E}\kern-.125emX}}
\begin{document}

\title{Effectively Testing System Configurations\\of Critical IoT Analytics Pipelines}

\author{
\IEEEauthorblockN{Morgan K. Geldenhuys\IEEEauthorrefmark{1},
Lauritz Thamsen\IEEEauthorrefmark{1},
Kain Kordian Gontarska\IEEEauthorrefmark{2},
Felix Lorenz\IEEEauthorrefmark{1} and
Odej Kao\IEEEauthorrefmark{1}}
\IEEEauthorblockA{\IEEEauthorrefmark{1}Technische Universit{\"a}t Berlin, Germany, \{firstname.lastname\}@tu-berlin.de}
\IEEEauthorblockA{\IEEEauthorrefmark{2}Hasso Plattner Institute, University of Potsdam, Germany, kordian.gontarska@hpi.de}
}

\maketitle
\copyrightnotice

\makeatletter

%\AddToShipoutPicture*{\small \sffamily\raisebox{1.2cm}{\hspace{1.8cm}978-1-7281-0858-2/19/\$31.00 © 2019 IEEE}}

%\def\ps@IEEEtitlepagestyle{%
  %\def\@oddfoot{\mycopyrightnotice}%
  %\def\@evenfoot{}%
%}
 %\def\mycopyrightnotice{%
   %{\footnotesize 978-1-7281-0858-2/19/\$31.00 © 2019 IEEE\hfill}
   %\gdef\mycopyrightnotice{}% just in case
 %}

\begin{abstract}

The emergence of the Internet of Things has seen the introduction of numerous connected devices used for the monitoring and control of even Critical Infrastructures. Distributed stream processing has become key to analyzing data generated by these connected devices and improving our ability to make decisions. However, optimizing these systems towards specific Quality of Service targets is a difficult and time-consuming task, due to the large-scale distributed systems involved, the existence of so many configuration parameters, and the inability to easily determine the impact of tuning these parameters.

In this paper we present an approach for the effective testing of system configurations for critical IoT analytics pipelines. We demonstrate our approach with a prototype that we called \emph{Timon} which is integrated with Kubernetes. This tool allows pipelines to be easily replicated in parallel and evaluated to determine the optimal configuration for specific applications. We demonstrate the usefulness of our approach by investigating different configurations of an exemplary geographically-based traffic monitoring application implemented in Apache Flink.

\end{abstract}


\section{Introduction}

The Internet of Things (IoT) is an important emerging technological paradigm whereby billions of ubiquitous sensor and actuator devices are connected to enable the development of applications across a wide number of domains. An increasing number of these applications are expected to perform in a capacity where the services they provide must meet certain minimum Quality of Service (QoS) requirements. This is especially relevant for applications used in the real-time monitoring and control of Critical Infrastructures, such as: human health-care, transportation systems, electrical generation, natural disaster prediction, and telecommunications, to name but a few \cite{GJF+16,JGM+14,CLC+15,LPS16}. 

As the number of Internet-connected devices increases year-on-year, so does the volume of data being produced. In order to process these large data streams, Distributed Stream Processing Frameworks (DSPF) such as Storm\cite{TTS+14}, and Flink\cite{CKE+15} allow for the deployment of analytics pipelines which utilize the processing power of a cluster of commodity nodes. Therefore, these frameworks are being utilized increasingly for the processing of IoT data streams \cite{FBK+16,SCS17,AGP17,JVR+18}. Applications developed within these systems are, in principle, required to operate indefinitely on an unbounded stream of continuous data in an environment where partial failures are to be expected as these applications scale. Consequently, DSPFs feature high availability modes, implement fault tolerance mechanisms by default, and expose a rich set of continually evolving features. The end result being that the way in which these systems are composed has a high level of complexity and number of configuration options. A quick scan of the official documentation reveals that Flink has over 300 options across 28 categories\footnote{Flink Configuration. URL: https://ci.apache.org/projects/flink/flink-docs-stable/ops/config.html}, and Spark\cite{ZCF+10} closer to 400 across 26 categories\footnote{Spark Configuration. URL: https://spark.apache.org/docs/latest/configuration}.

System configuration has an impact on performance and reliability. Yet, with the vast number of options available for tuning, i.e. framework settings, job parameters, resource selections, etc., the effects of which are not always well understood or straightforward to determine. That is, finding the best combination of resource selections and system configurations is difficult to estimate upfront both by experts and automatically by optimization tools as it is highly dependent on a number of key factors: \textit{the analytics application} which exhibits its own unique operational characteristics; \textit{the cluster environment} which is often not known before deployment and may vary over time, i.e. network topologies and physical hardware; and \textit{the data} which is variable based on the characteristics of the input data, loads from other applications, and ingestion rates. This is especially true in environments consisting of multiple connected distributed systems making up larger application architectures, such as: resource managers, messaging queues, distributed file systems, scalable databases, etc. At the same time, critical IoT applications typically have defined QoS requirements with regards to 
performance, reliability, etc, which a configuration should meet\cite{WNC17}.

Currently, the most common way of tuning configuration parameters is for it to be done manually by performance engineers, usually requiring several hours of investigation and testing \cite{AJP15}. These engineers require detailed knowledge of the specific DSPF itself and the cluster environment in order to find a system configuration that falls inline with the aforementioned QoS constraints. Approaches have been proposed at finding more precise and less time-consuming methods for the automatic tuning of DSPF parameters\cite{FGB15,JC16,BC17, TLW17,TWH19}. These typically focus on only a limited number of settings, while there are numerous points of configurations in practice with many dependencies between them. A solution is needed which is complementary to these existing performance modelling approaches, which provides an approach for gathering analytics data through testing and monitoring.

For this purpose, we propose an approach for the effective testing of system configurations for critical IoT analytics pipelines in realistic conditions. We implemented our approach using a prototype called \emph{Timon} which allows for the testing of multiple different versions of system configurations in parallel within an environment that behaves like production using real streaming data. In this way, operators can safely and efficiently experiment with potential system configurations to understand what impact these will have when used in production. 

The remainder of the paper is structured as follows: Section II discusses the related work with regards to configuring DSPFs, Section III presents a typical architecture for critical IoT analytics pipelines, Section IV presents our approach to configuration testing, Section V describes our evaluation where we present our experiments and findings, and Section VI discusses our findings with conclusions.

\section{Related Work}
\label{relatedwork}

%In this section we briefly discuss works closely related to our own: parameter tuning for DSPFs.

%\subsection{Parameter Tuning}

There exists a large body of work which addresses the problem of system configuration. A number of these approaches focus specifically on the tuning of parameters for DSPFs. This typically involves learning from analyzing actual executions or historical data, model specific aspects of the systems, and then adapt to actual conditions based on user requirements. We see our work as being orthogonal and complementary to these contributions in that Timon provides a testing and metric gathering environment within which these approaches could function. These approaches can be categorized as follows:

%\textit{Design of experiments}: A empirical approach where users try various combinations of parameters and analyze the results to find good configurations. Such an approach becomes less feasible the larger the multi-dimensional parameter space and therefore more advanced methods are required. 

\textit{Rule-based}: A gray-box heuristic approach where domain experts work with users to establish a rule-set which is used to recommend suitable configurations. Bilal et al. \cite{BC17} present an approach where users provide a parameter ranking in accordance with a priority level and specify whether an increase in parameter value has an overall positive impact on latency and throughput. This approach favors quickly finding a suitable configuration at the expense of optimality.
 
\textit{Model-based}: An approach which is concerned with conducting experiments on a chosen set of configurations to observe their performance. The results are used to train a statistical model for finding good configurations. Fischer et al. \cite{FGB15} and Trotter et al. \cite{TLW17} present an auto-tuning algorithm using Bayesian Optimization (BO)\cite{SSW+15} to achieve high throughput. Jamshidi et al. \cite{JC16} likewise proposes BO, however, it optimizes latency and leverages Gaussian Processes\cite{RN10} to continuously estimate the mean and confidence interval of a response variable at yet-to-be explored configurations. 

\textit{Search-based}: For this approach, an initial configuration is selected after which experiments are conducted sequentially. Each iteration uses the results of the previous to fit a statistical model which is used to select the next configuration. Evolutionary Search Algorithms are typically adopted for automatic parameter tuning. Trotter et al. \cite{TLW17} proposes a method using genetic algorithms (GA) to optimize throughput. Additionally, in a later paper Trotter et al. \cite{TWH19} use GA to optimize throughput using SVM classifiers to further refine its search of the configuration space. Bilal et al. \cite{BC17} proposes a hill-climbing algorithm based on Latin Hypercube Sampling\cite{MBC79} while taking both latency and throughput metrics into account.  

\textit{Learning-based}: An approach which use online learning techniques such as reinforcement learning to find the optimal configuration by reacting to feedback, i.e. metrics, from the DSPF at runtime \cite{VC18}. This approach can be combined with offline learning techniques to speed up convergence\cite{BRX09}. 

To the best of our knowledge, no approach exists which focuses on parameter tuning for critical IoT analytics pipelines executing in the production environment. For these applications it is essential to consider the time-dependant nature of IoT data streams when optimizing the performance of DSPFs.
 
%\subsection{Chaos Engineering}

%This methodology is about introducing controlled disruptions into a system, carefully studying the behaviors, identifying areas of weakness, and improving resiliency through automation\cite{BBD+16}. Resilience is often generalized as the ability of a system to tolerate failures while ensuring a specific Quality of Service. Chaos engineering, therefore, is a technique for meeting the resilience requirements. Although generally targeted towards web services, this technique provides us with a novel way of testing the fault tolerance behaviors of running stream processing analytics pipelines by injecting real-world inputs (e.g. transient network failures, surges in incoming requests, malformed data inputs, fail-stop failures).

\begin{figure*}
    \centering
    \includegraphics[width=0.8\textwidth]{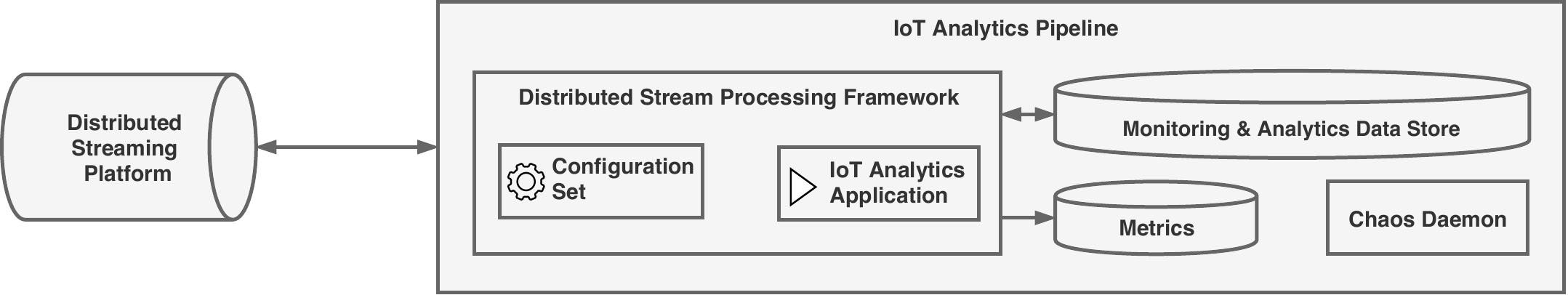}
    \caption{Typical IoT stream processing architecture.}
    \label{proposed}
\end{figure*}

\section{IoT Stream Processing Architecture}
\label{iotstreamprocessingarchitecture}

In this section we assume a typical architecture for the processing of IoT data streams, as depicted in Fig. \ref{proposed}. Here we see a number of systems which, when combined, provide a typical way of composing critical IoT analytics architectures. 

The \textit{distributed streaming platform} is where raw IoT data flows into the system from sensor devices and is stored in a messaging queue to await processing. This component is an implementation of the publisher/subscriber messaging pattern with Apache Kafka\cite{KNR11} being a distributed example of this. 

Subscribers such as the \textit{IoT analytics pipeline} register themselves with the distributed streaming platform and consume messages from targeted messaging queues when they becomes available. Additionally, messages that have already been processed and produce alarms or notifications, for instance, can be written back to a separate messaging queue for further consumption. The IoT analytics pipeline in turn consists of a number of inter-dependent systems. These systems include: 

\begin{itemize}
    \item \textit{Distributed Stream Processing Framework}: responsible for executing the \textit{IoT analytics application} with the current \textit{configuration set} in order to process messages read from the distributed streaming platform. Once processed, outputs are written back to the distributed streaming platform if necessary and the \textit{monitoring \& analytics data store} for archival. E.g. Apache Flink.
    
    \item \textit{Monitoring \& Analytics Data Store}: this scalable database warehouses all sanitized messages and outputs of the IoT analytics application. The archived data can be used for analysis over a longer period of time to detect trends and other anomalies. E.g. Apache Cassandra\cite{LM10}.
    
    \item \textit{Metrics}: it is important to monitor the health of the IoT analytics application being executed. For this purpose, most DSPFs like Flink and Spark generally have a metric system that allows for the gathering and export of internal metrics to external systems. These measurements should be stored in a time series database to be accessed from outside the cluster. E.g. InfluxDB\footnote{InfluxDB. URL: https://github.com/influxdata/influxdb}.
    
    \item \textit{Chaos Daemon}: an optional component, however, if deployed could provide the facility for using Chaos Engineering techniques to promote the development of resilient services\cite{BBD+16}. E.g. PowerfulSeal\footnote{PowerfulSeal. URL: https://github.com/bloomberg/powerfulseal}.
\end{itemize}

In such a setup, the configuration set and/or IoT analytics application could be designed to a standardized specification allowing them to be traded out for alternate versions without causing too much of a disruption to the overall environment.

\section{Approach}

\begin{figure*}
    \centering
    \includegraphics[width=\textwidth]{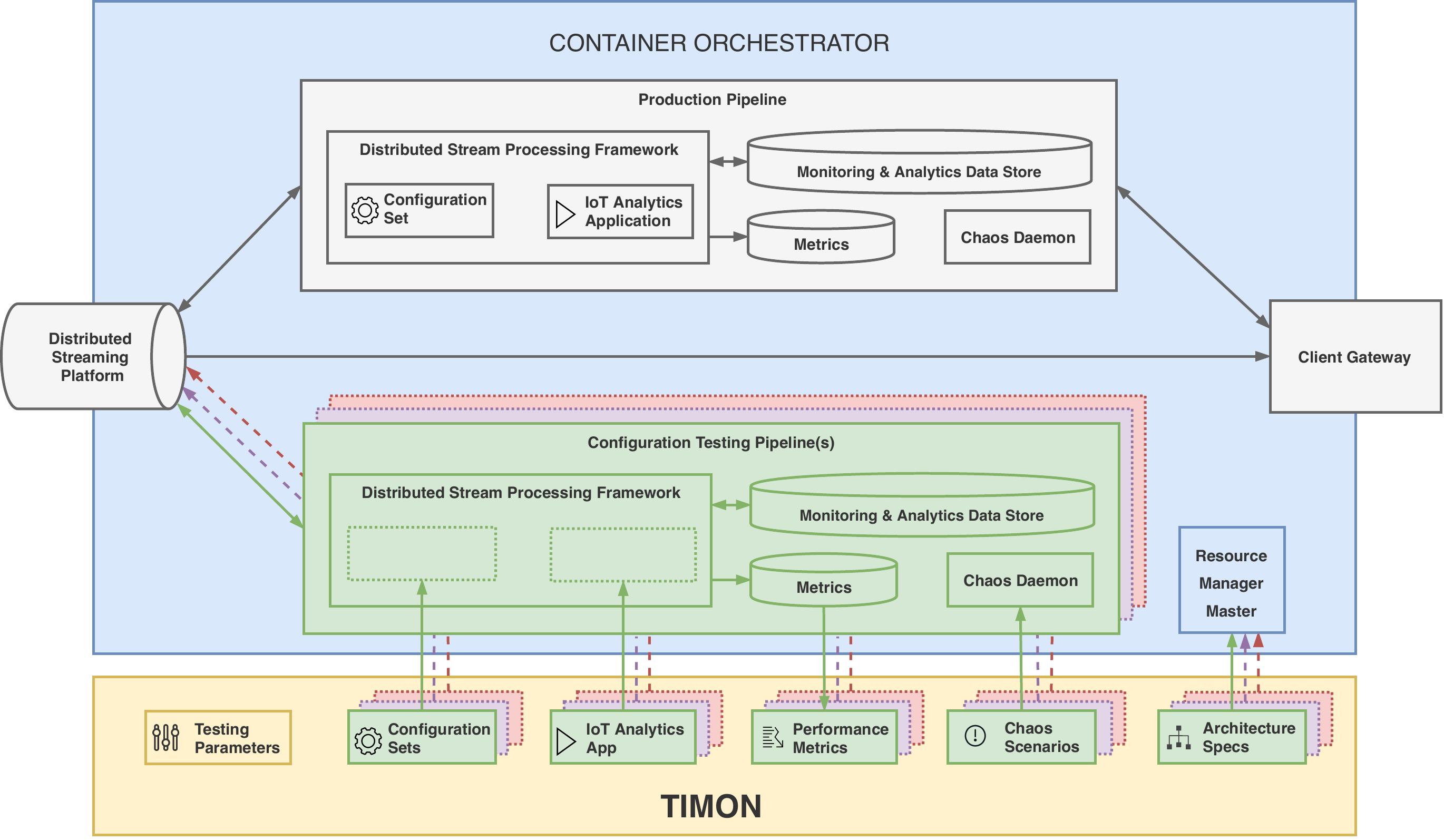}
    \caption{Overview of Timon and system dependencies.}
    \label{architecture}
\end{figure*}

In distributed stream processing, system configuration has a direct impact on performance and reliability. Yet, quantifying exactly how much of an impact is hard to ascertain. This is complicated by the existence of so many configuration parameters and the fact that no two stream processing applications with even a minor difference in operational characteristics is likely to share the same optimal setup. Additionally, if a more performant version of the configuration were to be found, migrating to this version in the production environment could cause significant disruptions. It is the goal of Timon to find solutions to these problems.

From a high level perspective, determining the best system configuration for any particular stream processing application can be found by comparing it to: the same application executing in the same environment while ingesting the same data but using alternate variations of the configuration set. For such a test, variants should be executed in parallel, metrics recorded over a specific time interval, and on conclusion, results compared to determine the best performer(s). This is the general idea behind Timon, to provide a testing system for the efficient comparison of alternative configurations in the production environment, i.e. testing with actually deployed systems, at scale, and with actual live data streams.

One of the key requirements of such a testing system would be the ability of an environment where alternate deployments can be quickly replicated. Moreover, these deployments would need to be isolated from each other in order to eliminate interference and provided with access to external services. For this purpose we make use of two key enabling technologies, i.e. OS-level virtualization and container orchestration. These technologies, when combined with Infrastructure-as-Code (IaC) processes, provide a mechanism for efficiently instantiating entire pipelines in parallel, assuming enough resources are available in the cluster to do so. 

Fig. \ref{architecture} provides an architectural overview of Timon and its dependencies. In this diagram we can see the virtual cluster environment where both the \textit{production pipeline} and shorter-lived \textit{configuration testing pipelines} exist and are managed by the container orchestrator. The flow of data is from left to right, from source, i.e. \textit{distributed streaming platform}, to sink, i.e. \textit{client gateway}. Each configuration testing pipeline is composed in the same way as the production pipeline and would be executing the same \textit{IoT analytics application}. Importantly, all configuration testing pipelines will also process the same input data as the production pipeline. When notifications and alarms produced by a configuration testing pipelines needs to be written back to the distributed streaming platform, they will each have their own unique messaging queues.

As part of the assumed \textit{IoT stream processing architecture} described in the previous section, each IoT analytics pipeline records metrics in a time series database. After all testing rounds have concluded, these metrics are collected, aggregated, and subsequently analyzed to determine if statistically significant effects were observed. If a better performing testing pipeline is found than the current production pipeline, then a strategy can be followed to replace it. This strategy first involves migrating all data not stored in the production pipeline to the new candidate pipeline, i.e. message queue and data store. Next, user traffic needs to be redirected towards the new data sources via the client gateway. In this way the client gateway can be thought of as a load balancer. Lastly, all redundant pipelines can then be safely decommissioned and resources recovered. It is important to note that the copying of archived data over the network can be an expensive operation both in terms of time and network resources. It is therefore prudent to follow a strategy which will minimize this impact. Container orchestrators such as Kubernetes\cite{VPK+15} offer a number of mechanisms for working with persisted data\footnote{Persistent Volumes. URL: https://kubernetes.io/docs/concepts/storage/persistent-volumes}.

Apart from performance, reliability testing is also important for understanding the behavior of distributed systems. The amount of things that can go wrong while a distributed system is running is enormous. This is mainly due to the distributed nature of all the components (which interact exclusively through direct message parsing). It is virtually impossible to predict every possible failure mode and then engineer solutions for all the edge cases. Instead, a more realistic approach would be to identify the weaknesses which cause these failures before they are triggered. This is where Chaos Engineering practices can be used to complement traditional testing approaches. Therefore, Timon provides the ability to optionally define \textit{failure scenarios} whereby failures are injected into the testing environment so that their impacts can be studied.

\section{Evaluation}

Now we demonstrate that using Timon is both practical and beneficial for distributed stream processing by presenting an experiment conducted to evaluate the impact of using different checkpoint intervals on the overall performance of the system. Ensuring that DSPFs are fault tolerant while running in the production environment is imperative, however, quantifying this impact is important when considering QoS.

\subsection{Prototype Implementation}

Timon is essentially a software client which interfaces directly with a container orchestrator to automatically manage the instantiation / destruction of container groups. These container groups compose the inter-dependent systems which in turn make up the individual analytics pipelines. For implementation, we use Docker\footnote{Docker. URL: https://docker.com} and Kubernetes. These technologies were selected because they provide an IaC approach for the management and provisioning of pipelines through machine-readable definition files. Additionally, Kubernetes provides a mechanism for isolating pipelines through the use of namespaces, thereby minimizing the possibility of interference. 

\subsection{IoT Data Stream \& Analytics Application}

For the purposes of this experiment, we created a simulation which mapped the streets and intersections of an area with one kilometer radius of central Berlin, Germany. In this area we generated a number of vehicles which travelled along various routes while providing an update message every 1 second. This update contained the: vehicle ID, vehicle type, current location, speed, and direction. We use a sinusoidal function to model traffic behaviors where the number of simultaneous vehicles is varied from a minimum of 25,000 to a maximum of 75,000 as a function of the time of day (\textit{t} in seconds). This was done to more closely resemble real traffic behaviors rather than a linear gradient, i.e. the number of vehicles gradually increases until a peak point (rush hour), before gradually decreasing again. Messages are submitted to an Apache Kafka cluster to await processing by the IoT analytics pipeline.

The live stream of traffic messages stored in Apache Kafka are consumed and analyzed using a DSPF. We use Apache Flink for our experiments as it has native support for fault-tolerant stream processing and is known for high performance and low latency \cite{CKE+15}. A Flink cluster implements a master-slave architecture which consists of two processes: the JobManager and the TaskManager. We developed an analytics application for the purpose of analysis and define the following task: Determine the total number of different vehicle types within the simulation area accumulated over a 5 minute window period. Results were outputted to an Apache Cassandra database. This task uses "group by" transformations where the stream was logically partitioned into disjointed partitions. All messages with the same key, therefore, were assigned to the same partition which allowed for a high level of parallelism. The long windowing period of 5 minutes results in a larger accumulating of state across the parallel tasks and therefore is a good fit for testing fault tolerance behaviors.

\subsection{Experimental Setup}

Our experimental setup consists of a 3 node Apache Kafka cluster and a 30 node Kubernetes cluster with HDFS\cite{SKR+10}. Node specifications are shown in Table \ref{clusterspecs}.

\begin{table}[ht]
\centering
\caption{Cluster specifications}
\begin{tabular}[t]{rp{0.65\linewidth}}
\toprule
Resource&Details\\
\midrule
OS&Ubuntu 18.04.3\\
CPU&Quadcore Intel Xeon CPU E3-1230 V2 3.30GHz\\
Memory&16 GB RAM\\
Storage&3TB RAID0 (3x1TB disks, linux software RAID)\\
Network&1 GBit Ethernet NIC\\
Software&Java v1.8, Apache Flink v1.9.0, Apache Kafka v2.3.0, Apache ZooKeeper v3.5.5, Docker v18.06, Kubernetes v1.15.3, Apache HDFS V2.8.3, Apache Cassandra v3.11.4, InfluxDB v1.6.4 \\
\bottomrule
\end{tabular}
\label{clusterspecs}
\end{table}%

We limit our configuration sets to vary a single variable, i.e. checkpoint interval, and choose 3 different variations representing short (\textit{1000ms}), medium (\textit{20,000ms}), and long (\textit{120,000ms}) intervals. Using Timon, we created 3 corresponding configuration testing pipelines in Kubernetes, each composed of: 

\begin{itemize}
  \item an Apache Flink (High Availability) cluster of 11 instances (1 JobManagers and 10 TaskManagers);
  \item an Apache ZooKeeper cluster of 3 instances for distributed coordination; 
  \item an Apache Cassandra cluster of 3 instances for archival of processed data; and
  \item a single InfluxDB time series database instance for collection of performance measurements.
\end{itemize}

%We used Apache Flink to process the IoT traffic data stream. It has native support for fault-tolerant stream processing and is known for high performance and low latency \cite{CKE+15}. 

We define four key indicators to measure the performance of each configuration set. These are: \textit{end-to-end latency}, in DSPFs, is the time difference between the moment a message is produced at the source task and the moment the tuple is produced at the output; \textit{input throughput}, measured in messages / sec, is the cumulative frequency at which messages enter the source tasks of the dataflow, and; \textit{CPU utilization} and \textit{heap memory utilization} as a percentage.

The total time to provision pipelines for each experimental testing round averaged 330 seconds. There were 5 rounds of testing conducted where metrics were recorded over 6 hours with an increasing input throughput of 25,000 to 75,000 messages per second. 

%It is important to note that the increase in input latency was not linear and followed the sinusoidal curve mentioned earlier in this section. 

\subsection{Experimental Results}

In the experiments, during the user-defined time interval, metrics were recorded and saved to a time-series database. After all testing rounds were concluded, Timon automatically retrieved these metrics, aggregated them, and analysis was performed. Parallelism for the dataflow job was set to eight. This resulted in one sink operator executing for each of the eight active TaskManagers. Latency, therefore, is recorded at each sink operator separately. In order to address the observed variance in the performance metrics, the median latency value from all sink operators was chosen for each timestamp. The same was applied to the TaskManagers for CPU and memory utilization. Furthermore, the experiment was run for a total of five testing rounds over the same time interval, i.e. time of day. Again, the median values for each time step were chosen to be the expected values. To further remove noise from the diagrams, exponential weighted moving average windows with a span of 1000 seconds were applied to the averaged metrics.

%In order to address the observed variance in our performance metrics, we filtered outliers in order to provide more appropriate expected values. The These results were then averaged 

%In order to aggregate these values and overcome the observed variance in the performance metrics, outliers were excluded to provide more appropriate expected values. The remaining measurements were averaged

%Latency is measured for each TaskManager separately. 

%As is often observed in distributed environments, performance metrics exhibited a high level of variance due to the distributed nature of these systems [TODO ref]. To address this, we filtered outliers in order to provide more appropriate expected values and subsequently averaged the measured metrics.

%Latency is measured for each TaskManager separately. The median latency values from all TaskManagers are chosen for each timestamp. The same is applied to the TaskManagers' CPU and memory utilization. Furthermore, the experiment has been run for 5 rounds over the same time interval, i.e. time of day. Again, the median values for each time step were chosen to be the expected values. To further remove noise from the diagrams, exponential weighted moving average windows with a span of 1000 were applied to the averaged metrics.

\begin{figure}[htb]
    \centering
    \includegraphics[width=0.99\linewidth]{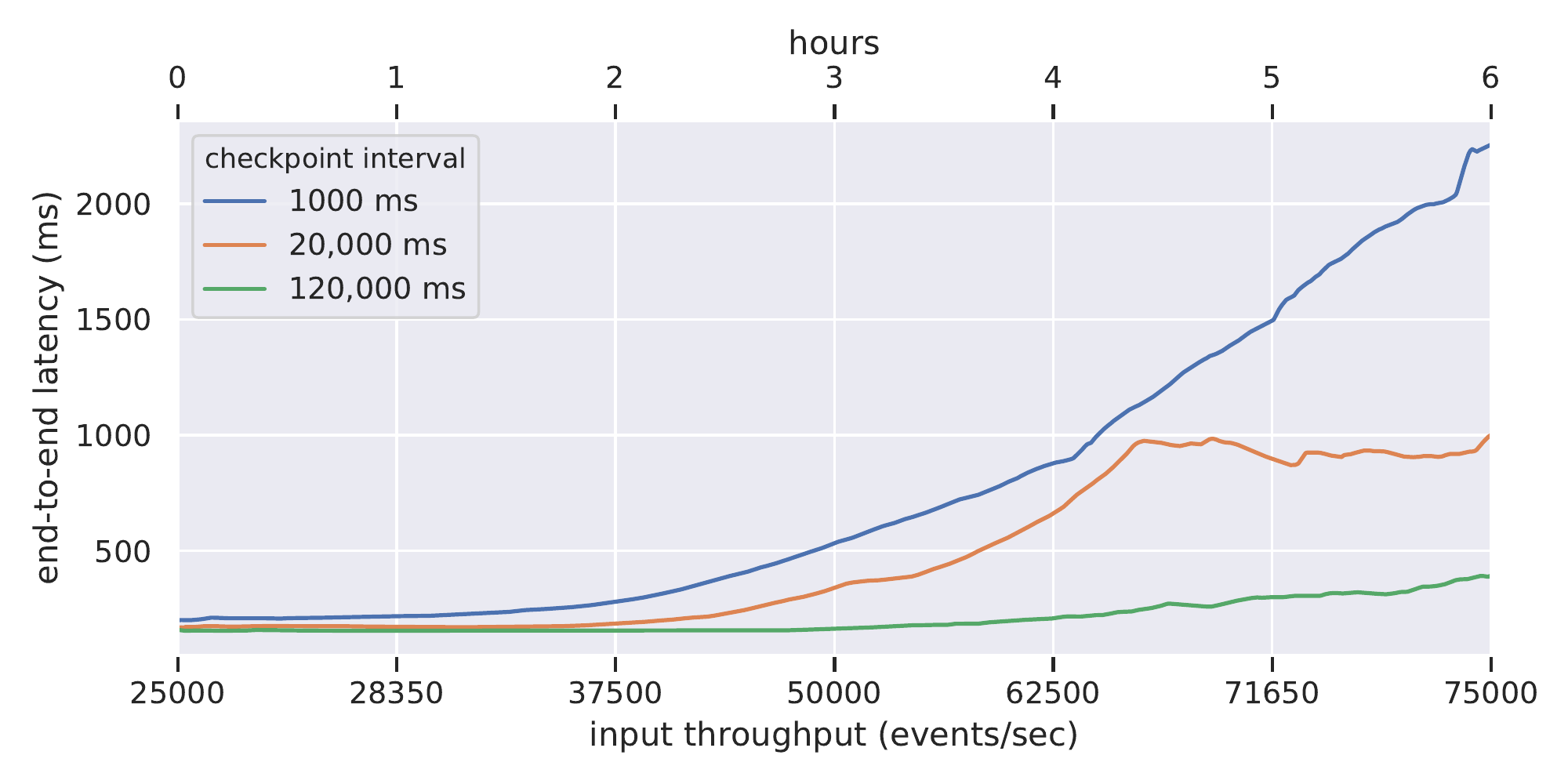}
    \caption{End-to-end latencies.}
    \label{latency}
\end{figure}

Fig. \ref{latency} shows the average latencies as measured at the sink operator across all TaskManagers. Here we can clearly see how, as input throughput increases, performance deteriorates across all configurations. Additionally, latencies decrease more drastically the shorter the checkpoint interval as input throughput increases. It is visible there is a trade-off between QoS requirements, i.e. the maximum amount of time before a message should be processed, and the recovery time should a failure occur, i.e. the time for the system to go to a state where all messages are processed up until the failure.%point before the failure was experienced.

\begin{figure}[htb]
    \centering
    \includegraphics[width=0.98\linewidth]{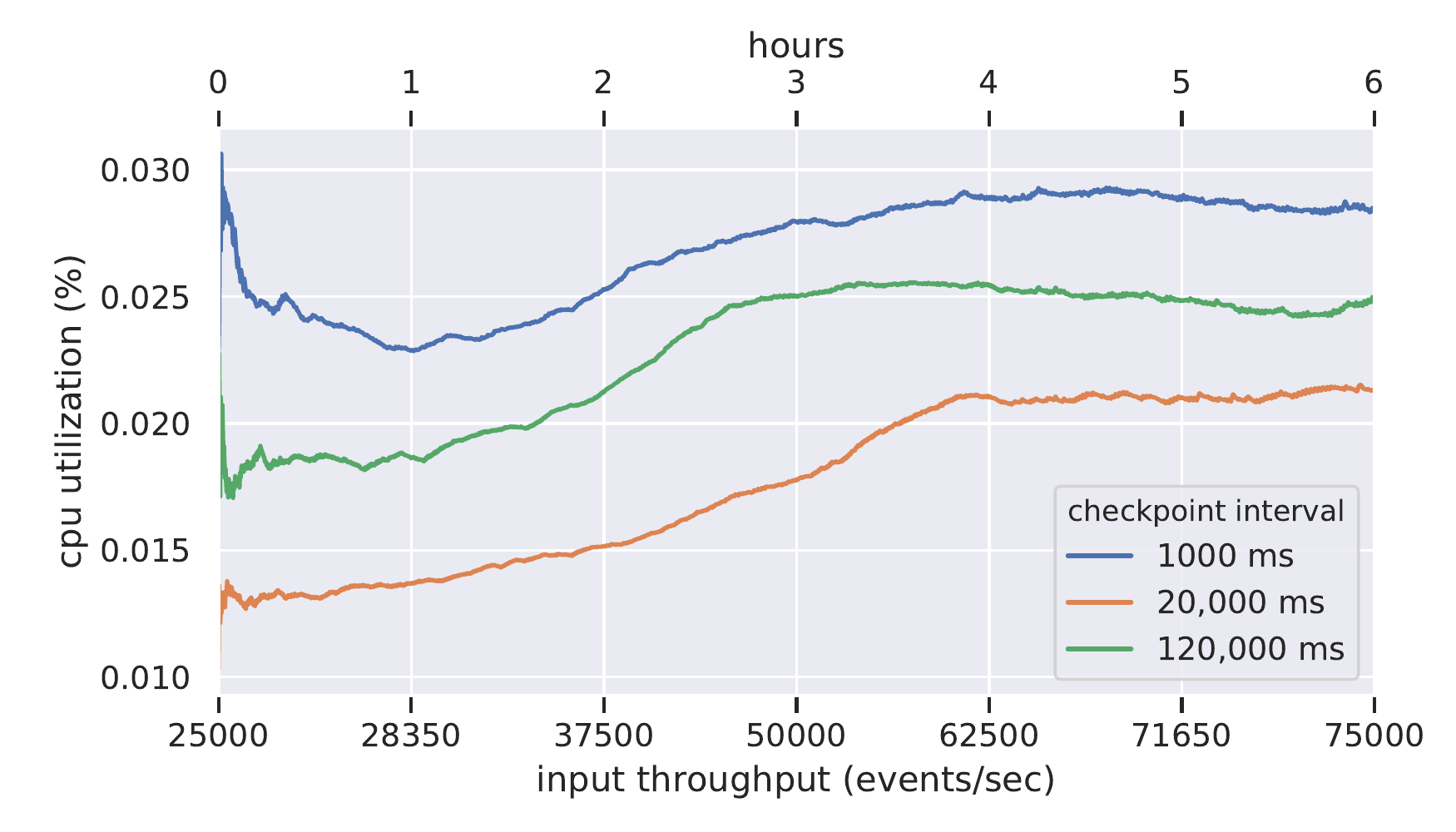}
    \caption{TaskManager CPU utilization.}
    \label{cpu}
\end{figure}

\begin{figure}[htb]
    \centering
    \includegraphics[width=0.98\linewidth]{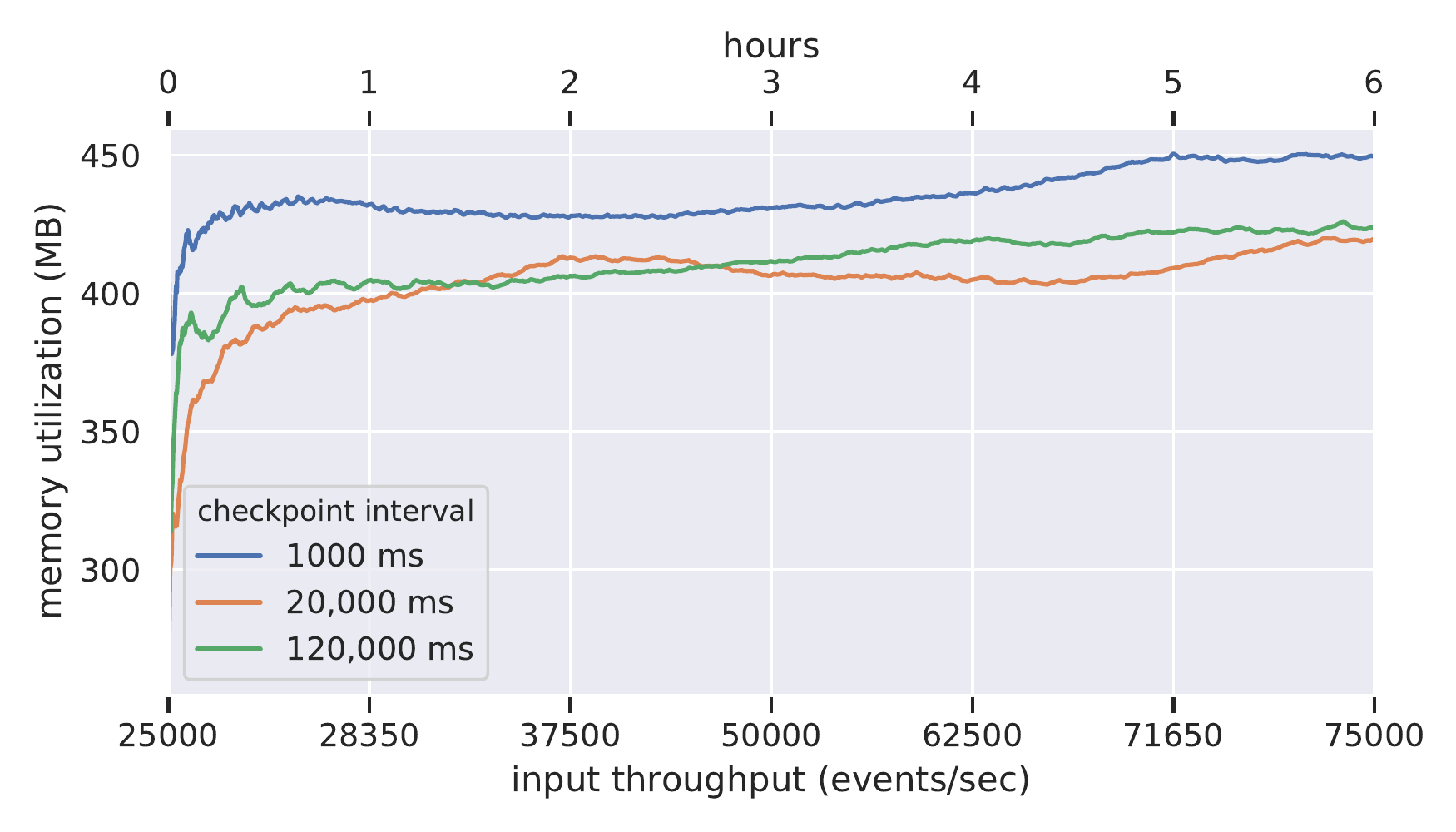}
    \caption{TaskManager memory utilization.}
    \label{memory}
\end{figure}

Fig. \ref{cpu} and \ref{memory} show the average resource utilization for CPU and memory across all TaskManagers. Here we can see how CPU and memory usage increases as input throughput increases, however, in both cases utilization is low and there is no need to provision more resources through configuration. The default setting for memory assigned to each TaskManager is 1 GB. Analysis of resource utilization for the JobManagers was likewise performed. As is to be expected, CPU and memory utilization was lower than the TaskManagers while following the same trend of the shorter the checkpoint interval, the more resources were consumed and this increases as input throughput increases.

%[TODO what to do with these values?]

%memory, JobManager: \\
%config-1: min 49.627112MB max 384.411152MB \\
%config-2: min 43.800808MB max 338.321568MB \\
%config-3: min 45.303952MB max 338.520864MB \\

%memory, TaskManager: \\
%config-1: min 161.351208 MB max 711.7604MB \\
%config-2: min 77.913284MB max 683.432188MB \\
%config-3: min 89.432344MB max 670.774888MB \\

%cpu, JobManager: \\
%config-1: min 0.0\% max 27.27\% \\
%config-2: min 0.1\% max 12.5\% \\
%config-3: min 0.09\% max 3.92\% \\

%cpu, TaskManager: \\
%config-1: min 0.09\% max 12.71\% \\
%config-2: min 0.13\% max 12.22\% \\
%config-3: min 0.14\% max 12.73\% \\

\section{Conclusion}

This paper presented an approach which allows for the effective testing of system configuration of critical IoT analytics pipelines in realistic conditions. For this, we assume a typical distributed architecture for critical IoT analytics pipelines and utilize containerization as well as container-orchestration in order to replicate instances of this architecture in parallel, each with their own configuration set. We showed how using such a testing approach in the production environment can capture the runtime behaviors of stream processing applications in order to investigate the individual performance of each configuration set. This was done by aggregating chosen metrics recorded over a defined number of testing rounds and then comparing them. Ultimately, the choice of which configuration set is the best performer should always consider pre-defined QoS requirements. 

In the future, we would like to expand upon our approach in two ways. Firstly, we want to conducting experiments with failure scenarios and including critical IoT analytics applications from different domains in addition to smart city. Secondly, we want to research flexible methods for automatic parameter tuning and selection of optimal performing configurations. Nevertheless, this approach has already proven to be a helpful testing method and a usable tool.

%This architecture consisted of: a streaming platform to store pre- and post-processed messages, a DSPF to execute the analytics application together with a customized version of the system configurations, a scalable data store for archival, and a time-series database for the recording of metrics.

%For the implementation of our prototype, we used Docker and Kubernetes for container orchestration, and Apache Flink as the DSPF to execute our IoT traffic processing application.

% adding to the number of DSPFs which were used for testing, i.e. including Apache Spark in addition to Apache Flink;

\section*{Acknowledgments} \label{sec:ACKNOWLEDGMENTS}

This work has been supported through grants by the German Ministry for Education and Research (BMBF) as Berlin Big Data Center BBDC2 (funding mark 01IS18025A).
% TODO: add WaterGridSense!

\bibliographystyle{IEEEtran}
\bibliography{bib}

\begin{thebibliography}{10}
\providecommand{\url}[1]{#1}
\csname url@samestyle\endcsname
\providecommand{\newblock}{\relax}
\providecommand{\bibinfo}[2]{#2}
\providecommand{\BIBentrySTDinterwordspacing}{\spaceskip=0pt\relax}
\providecommand{\BIBentryALTinterwordstretchfactor}{4}
\providecommand{\BIBentryALTinterwordspacing}{\spaceskip=\fontdimen2\font plus
\BIBentryALTinterwordstretchfactor\fontdimen3\font minus
  \fontdimen4\font\relax}
\providecommand{\BIBforeignlanguage}[2]{{%
\expandafter\ifx\csname l@#1\endcsname\relax
\typeout{** WARNING: IEEEtran.bst: No hyphenation pattern has been}%
\typeout{** loaded for the language `#1'. Using the pattern for}%
\typeout{** the default language instead.}%
\else
\language=\csname l@#1\endcsname
\fi
#2}}
\providecommand{\BIBdecl}{\relax}
\BIBdecl

\bibitem{GJF+16}
D.~Georgakopoulos, P.~Jayaraman, M.~Fazia, M.~Villari, and R.~Ranjan,
  ``Internet of things and edge cloud computing roadmap for manufacturing,''
  \emph{IEEE Cloud Computing}, vol.~3, pp. 66--73, 2016.

\bibitem{JGM+14}
J.~Jin, J.~Gubbi, S.~Marusic, and M.~Palaniswami, ``An information framework
  for creating a smart city through internet of things,'' \emph{IEEE Internet
  of Things Journal}, vol.~1, pp. 112--121, 2014.

\bibitem{CLC+15}
B.~Cheng, S.~Longo, F.~Cirillo, M.~Bauer, and E.~Kovacs, ``Building a big data
  platform for smart cities: Experience and lessons from santander,''
  \emph{2015 IEEE International Congress on Big Data}, pp. 592--599, 2015.

\bibitem{LPS16}
M.~Lom, O.~Pribyl, and M.~Svitek, ``Industry 4.0 as a part of smart cities,''
  \emph{2016 Smart Cities Symposium Prague (SCSP)}, pp. 1--6, 2016.

\bibitem{TTS+14}
A.~Toshniwal, S.~Taneja, A.~Shukla, K.~Ramasamy, J.~M. Patel, S.~Kulkarni,
  J.~Jackson, K.~Gade, M.~Fu, J.~Donham, N.~A. Bhagat, S.~Mittal, and
  D.~Ryaboy, ``Storm@twitter,'' \emph{Proceedings of the 2014 ACM SIGMOD
  International Conference on Management of Data}, 2014.

\bibitem{CKE+15}
P.~Carbone, A.~Katsifodimos, S.~Ewen, V.~Markl, S.~Haridi, and K.~Tzoumas,
  ``Apache flink: Stream and batch processing in a single engine,'' \emph{IEEE
  Data Eng. Bull.}, vol.~38, pp. 28--38, 2015.

\bibitem{FBK+16}
G.~Morales, A.~Bifet, L.~Khan, J.~Gama, and W.~Fan, ``Iot big data stream
  mining,'' \emph{Proceedings of the 22nd ACM SIGKDD International Conference
  on Knowledge Discovery and Data Mining}, 2016.

\bibitem{SCS17}
A.~Shukla, S.~Chaturvedi, and Y.~Simmhan, ``Riotbench: An iot benchmark for
  distributed stream processing systems,'' \emph{Concurrency and Computation:
  Practice and Experience}, vol.~29, 2017.

\bibitem{AGP17}
S.~Amini, I.~Gerostathopoulos, and C.~Prehofer, ``Big data analytics
  architecture for real-time traffic control,'' \emph{2017 5th IEEE
  International Conference on Models and Technologies for Intelligent
  Transportation Systems (MT-ITS)}, pp. 710--715, 2017.

\bibitem{JVR+18}
G.~Jansen, I.~Verbitskiy, T.~Renner, and L.~Thamsen, ``Scheduling stream
  processing tasks on geo-distributed heterogeneous resources,'' \emph{2018
  IEEE International Conference on Big Data (Big Data)}, pp. 5159--5164, 2018.

\bibitem{ZCF+10}
M.~Zaharia, M.~Chowdhury, M.~Franklin, S.~Shenker, and I.~Stoica, ``Spark:
  Cluster computing with working sets,'' in \emph{HotCloud}, 2010.

\bibitem{WNC17}
G.~White, V.~Nallur, and S.~Clarke, ``Quality of service approaches in iot: A
  systematic mapping,'' \emph{J. Syst. Softw.}, vol. 132, pp. 186--203, 2017.

\bibitem{AJP15}
S.~Allen, M.~Jankowski, and P.~Pathirana, ``Storm applied: Strategies for
  real-time event processing,'' 2015.

\bibitem{FGB15}
L.~Fischer, S.~Gao, and A.~Bernstein, ``Machines tuning machines: Configuring
  distributed stream processors with bayesian optimization,'' \emph{2015 IEEE
  International Conference on Cluster Computing}, pp. 22--31, 2015.

\bibitem{JC16}
P.~Jamshidi and G.~Casale, ``An uncertainty-aware approach to optimal
  configuration of stream processing systems,'' \emph{2016 IEEE 24th
  International Symposium on Modeling, Analysis and Simulation of Computer and
  Telecommunication Systems (MASCOTS)}, pp. 39--48, 2016.

\bibitem{BC17}
M.~Bilal and M.~Canini, ``Towards automatic parameter tuning of stream
  processing systems,'' \emph{Proceedings of the 2017 Symposium on Cloud
  Computing}, 2017.

\bibitem{TLW17}
M.~Trotter, G.~Liu, and T.~Wood, ``Into the storm: Descrying optimal
  configurations using genetic algorithms and bayesian optimization,''
  \emph{2017 IEEE 2nd International Workshops on Foundations and Applications
  of Self* Systems (FAS*W)}, pp. 175--180, 2017.

\bibitem{TWH19}
M.~Trotter, T.~Wood, and J.~Hwang, ``Forecasting a storm: Divining optimal
  configurations using genetic algorithms and supervised learning,'' \emph{2019
  IEEE International Conference on Autonomic Computing (ICAC)}, pp. 136--146,
  2019.

\bibitem{SSW+15}
B.~Shahriari, K.~Swersky, Z.~Wang, R.~Adams, and N.~D. Freitas, ``Taking the
  human out of the loop: A review of bayesian optimization,'' \emph{Proceedings
  of the IEEE}, vol. 104, pp. 148--175, 2016.

\bibitem{RN10}
C.~Rasmussen and H.~Nickisch, ``Gaussian processes for machine learning (gpml)
  toolbox,'' \emph{J. Mach. Learn. Res.}, vol.~11, pp. 3011--3015, 2010.

\bibitem{MBC79}
M.~McKay, R.~Beckman, and W.~Conover, ``A comparison of three methods for
  selecting values of input variables in the analysis of output from a computer
  code,'' \emph{Technometrics}, vol.~42, pp. 55 -- 61, 2000.

\bibitem{VC18}
L.~Vaquero and F.~Cuadrado, ``Auto-tuning distributed stream processing systems
  using reinforcement learning,'' \emph{ArXiv}, vol. abs/1809.05495, 2018.

\bibitem{BRX09}
X.~Bu, J.~Rao, and C.~Xu, ``A reinforcement learning approach to online web
  systems auto-configuration,'' \emph{2009 29th IEEE International Conference
  on Distributed Computing Systems}, pp. 2--11, 2009.

\bibitem{KNR11}
J.~Kreps, ``Kafka : a distributed messaging system for log processing,'' 2011.

\bibitem{LM10}
A.~Lakshman and P.~Malik, ``Cassandra: a decentralized structured storage
  system,'' \emph{ACM SIGOPS Oper. Syst. Rev.}, vol.~44, pp. 35--40, 2010.

\bibitem{BBD+16}
A.~Basiri, N.~Behnam, R.~D. Rooij, L.~Hochstein, L.~Kosewski, J.~Reynolds, and
  C.~Rosenthal, ``Chaos engineering,'' \emph{IEEE Software}, vol.~33, pp.
  35--41, 2016.

\bibitem{VPK+15}
A.~Verma, L.~Pedrosa, M.~Korupolu, D.~Oppenheimer, E.~Tune, and J.~Wilkes,
  ``Large-scale cluster management at google with borg,'' \emph{Proceedings of
  the Tenth European Conference on Computer Systems}, 2015.

\bibitem{SKR+10}
K.~Shvachko, H.~Kuang, S.~Radia, and R.~Chansler, ``The hadoop distributed file
  system,'' \emph{2010 IEEE 26th Symposium on Mass Storage Systems and
  Technologies (MSST)}, pp. 1--10, 2010.

\end{thebibliography}

\end{document}